\documentclass[12pt]{article}
\usepackage{epsf}
\newcommand{\be}{\begin{equation}}
\newcommand{\ee}{\end{equation}}
\newcommand{\bea}{\begin{eqnarray}}
\newcommand{\eea}{\end{eqnarray}}
\newcommand{\nn}{\nonumber}

\begin{document}

\begin{titlepage}
\begin{center}
{\Large \bf William I. Fine Theoretical Physics Institute \\
University of Minnesota \\}
\end{center}
\vspace{0.2in}
\begin{flushright}
FTPI-MINN-07/22 \\
UMN-TH-2611/07 \\
July 2007 \\
\end{flushright}
\vspace{0.3in}
\begin{center}
{\Large \bf Remarks on the amplitude of the decay $\Upsilon(3S) \to \Upsilon(1S)
\, \pi
\pi$
\\}
\vspace{0.2in}
{\bf S. Dubynskiy \\}
School of Physics and Astronomy, University of Minnesota, \\ Minneapolis, MN
55455 \\
and \\
{\bf M.B. Voloshin  \\ }
William I. Fine Theoretical Physics Institute, University of
Minnesota,\\ Minneapolis, MN 55455 \\
and \\
Institute of Theoretical and Experimental Physics, Moscow, 117218
\\[0.2in]
\end{center}

\begin{abstract}

We point out some properties of the amplitude of the dipion transition
$\Upsilon(3S) \to \Upsilon(1S) \, \pi \pi$ in relation to the recently reported
results of a CLEO analysis of form factors in this amplitude. We find that the
reported significant complex phase between two of the form factors under the
assumption that the third form factor is zero, is not consistent with the
picture where the phase shifts arise due to the final state interaction in the
$\pi \pi$ channel. It is also shown that in an analysis that uses no information
on the polarization of both the initial and the final $\Upsilon$ resonances it
is impossible in principle to determine all the relevant terms. We suggest that
a study of a simple correlation between the direction of the total momentum of
the two pions and the axis of the initial beams is sufficient to resolve the
ambiguity in the fit for the form factors.

\end{abstract}

\end{titlepage}

The transitions between states of heavy quarkonium with emission of two pions
present a case study in both the internal works of heavy quarkonia and the low-
energy dynamics of light mesons in QCD. The general constraints on the
amplitudes of such transitions with soft pions follow from the chiral
algebra\cite{bc,mv0}. The specifics of heavy quarkonium is revealed in that any
such transition can be viewed, in a way, as a two-stage process: the quarkonium
transition produces gluon field, which in turn creates the light mesons.
Considering quarkonium as a compact object in the scale of typical energy in the
transition, one can apply the multipole expansion in QCD for interaction of the
heavy quark system with the glue field\cite{gottfried}, while the creation of
the two pions by this field can be described with the help of low-energy
theorems in QCD\cite{vz,ns}. It is well known that some of the observed $\pi
\pi$ transitions in charmonium, namely, $\psi(2S) \to J/\psi \, \pi
\pi,~\Upsilon(2S) \to \Upsilon(1S) \, \pi \pi$, and $\Upsilon(3S) \to
\Upsilon(2S) \, \pi \pi$ display the behavior, in particular the spectrum of the
invariant mass of the dipion, that is expected under the simplest assumptions
about the corresponding heavy quarkonium multipoles. On the other hand, the
transition $\Upsilon(3S) \to \Upsilon(1S) \, \pi \pi$ is long known to defy such
straightforward predictions by displaying a peculiar double-peaked spectrum of
$m_{\pi\pi}$\,\footnote{It is interesting to note that the data on the recently
observed two-pion transitions from $\Upsilon(4S)$ indicate that the transition
$\Upsilon(4S) \to \Upsilon(1S) \, \pi \pi$ has a well-behaved $m_{\pi \pi}$
spectrum\cite{belle,babar}, while the spectrum in the decay$\Upsilon(4S) \to
\Upsilon(2S)
\, \pi \pi$  possibly resembles that in $\Upsilon(3S) \to \Upsilon(1S) \, \pi
\pi$\,\cite{babar}. It is however unclear at present to what extent effects
of $B {\overline B}$ meson pairs can contribute to hadronic transitions from
$\Upsilon(4S)$.}.

Given such behavior, a further understanding of the anomalous behavior in the
transition $\Upsilon(3S) \to \Upsilon(1S) \, \pi \pi$ would likely greatly
benefit from an input from experiment providing more details about the structure
of the decay amplitude. A study of this decay based on a large data sample has
been recently reported by CLEO\cite{cleo}. In this analysis they use the
parametrization of the amplitude\cite{bc} motivated by the soft pion limit:
\bea
{\cal M} &=&
\left [ A \, (q^2-2 m_\pi^2) + \lambda \, m_\pi^2 \right ] \, \left ( \epsilon_1
\cdot \epsilon_2 \right ) + B \, E_1 E_2 \, \left ( \epsilon_1 \cdot \epsilon_2
\right ) \nn \\ &+& C \, \left [ (p_1 \cdot \epsilon_1) (p_2 \cdot \epsilon_2)+
(p_2 \cdot \epsilon_1) (p_1 \cdot \epsilon_2) \right ]~,
\label{abc}
\eea
where $\epsilon_1^{\mu}$ and $\epsilon_2^\mu$ are the polarization amplitudes of
the initial and the final vector resonances, $p_1$ and $p_2$ are the 4-momenta
of the two pions, $E_1$ and $E_2$ are their energies in the rest frame of the
initial state, and $q=p_1+p_2$ is the total 4-momentum of the dipion, so that
$q^2=m_{\pi \pi}^2$. Finally, $A$, $B$, $C$, and $\lambda$ are the form factors,
which
should be considered constant in the soft pion limit, but are generally
functions of kinematic variables beyond this limit. The term of order
$m_\pi^2$ proportional to $\lambda$ is the `$\sigma$ term'\cite{bc} and it is
fixed at zero in the fits of Ref.\cite{cleo}.  The spectrum of $m_{\pi\pi}$ and
the angular distribution in the angle $\theta_X$ between the direction of motion
of the two pions in their center of mass frame and the direction of the vector
${\vec q}$ are then used to fit the values of the ratios $b = B/A$ and $c =
C/A$. The results of this analysis suggest that the ratio $b$ has a
significant imaginary part: if $c$ is fixed as $c=0$, the fit for $b$
yields\cite{cleo} Re$(b) = -2.523 \pm 0.031$ and Im$(b)=\pm 1.189 \pm 0.051$
(obviously, the sign of the imaginary part cannot be determined in such fit). If
$c$ is allowed to float, it is found that $|c| < 1.09$ at 90\% C.L., and a
certain correlation between $|b|$ and $|c|$ is noted in the fit.

The purpose of our present paper is to elaborate on several points related to
the description of the decay amplitude and on the possibilities of studying the
details of such description from data. Specifically, we point out that using
only the phase-space distribution over $m_{\pi \pi}$ and $\theta_X$ and no
information on polarization of either the initial  or final $\Upsilon$
resonance, it is impossible in principle to unambiguously determine the ratia
$b$ and $c$, unless $c$ is exactly equal to zero. In other words, if $b$ and $c$
are allowed to be complex, there is a continuous set of transformations of $b$
and $c$ that does not change the distribution of the decay rate over $m_{\pi
\pi}$ and $\theta_X$. This degeneracy becomes a discrete two-fold ambiguity if
$b$ and $c$ are constrained to be real.

Furthermore, the imaginary part of the amplitude is related by the unitarity
condition to rescattering of the decay products. Given an apparent absence of a
strong rescattering in the exotic channel with the quantum numbers of $\Upsilon
\, \pi$, any complex phase behavior in the considered amplitude can arise only
from $\pi \pi$ rescattering, which is reduced to the relative phase between the
$S$ and $D$ wave $\pi \pi$ scattering phases in the isoscalar $I=0$ state. The
amplitudes for production of these partial waves combine in $b$ with
coefficients having different dependence on $m_{\pi \pi}$. For this reason any
complex phase of $b$ cannot be approximated as constant. Moreover, the
data\cite{cleo} suggest that the production of the $D$ wave is very small, so
that the interference between the $S$ and $D$ waves possible in parts of the
phase space (but not e.g. in the $m_{\pi \pi}$ spectrum) cannot be significant,
and effectively the relevant ratio $b$ is essentially real. Thus we believe that
the result\cite{cleo} claiming a sizable complex phase of $b$ is very likely
to be modified by further analyses.

Finally, a better way of independently determining the coefficients $b$ and $c$,
which does not suffer from the mentioned ambiguity, should use at least some
basic information about the polarization of either the initial or the final
$\Upsilon$ resonances. We point out that for the  $e^+e^-$ annihilation setting
the apparently simplest correlation sensitive to the polarization-dependent
amplitude is that between the direction of the initial beams and the direction
of the motion of the dipion, i.e. of the vector ${\vec q}$.

For consideration of the effects of different terms of the amplitude in the
observable phase space distribution and also for evaluating the significance of
the $\pi \pi$ rescattering it is helpful to write the decay amplitude as
a sum of partial waves\cite{mv1}:
\be
{\cal M}= S \, (\epsilon_1 \cdot \epsilon_2) + D_1 \, \ell_{\mu \nu} \, {P^\mu P
^\nu \over P^2} \, (\epsilon_1 \cdot \epsilon_2) + D_2\, q_\mu \, q_\nu \,
\epsilon^{\mu \nu}+ D_3 \, \ell_{\mu \nu} \, \epsilon^{\mu \nu}~.
\label{sddd}
\ee
In this expression $P$ is the 4-momentum of the initial resonance. The tensor
$\ell_{\mu
\nu}$ corresponds to a $D$-wave spatial tensor made out of momenta of the pions
in their $c.m.$ frame. Namely, using the notation $r=p_1-p_2$, this tensor is
defined as\cite{ns}
\be
\ell_{\mu \nu}=r_\mu r_\nu  + {1 \over 3}\, \left ( 1- {4 m_\pi^2 \over q^2}
\right ) (q^2 \, g_{\mu \nu} - q_\mu q _\nu)~.
\label{lmn}
\ee
Finally, $\epsilon^{\mu \nu}$ stands for the spin-2 tensor made from the
polarization amplitudes of the resonances
\be
\epsilon^{\mu \nu}=\epsilon_1^\mu \epsilon_2^\nu +  \epsilon_1^\nu
\epsilon_2^\mu +{2 \over 3} \, (\epsilon_1 \cdot \epsilon_2) \, \left ( {P^\mu P
^\nu \over P^2} - g_{\mu \nu} \right )~.
\label{emn}
\ee

The terms in the expression (\ref{sddd}) describe an $S$ wave and three possible
types of $D$-wave motion: the term with $D_1$ corresponds to a $D$ wave in the
c.m. system of the two pions correlated with the overall motion of the dipion in
the rest frame of the initial state, the $D_2$ term describes the $D$-wave
motion of the dipion as a whole, correlated with the spins of the $\Upsilon$
resonances, and finally, the $D_3$ term corresponds to the correlation between
the spins and the $D$-wave motion in the c.m. frame of the dipion. One can also
notice that the $S$ and $D_1$ terms contain an overall spin-0
combination of the quarkonium polarizations, so that there is no interference
between these two terms and those with $D_2$ and $D_3$, if no polarization
information in the rate is used. In particular, the distribution of the rate
studied in Ref.\cite{cleo} can be written as
\be
{d\Gamma\over d\cos\theta_X \, d q} \propto {\overline {|{\cal M}|^2_X}} \,
\sqrt{q_0^2 - q^2} \, \sqrt{q^2-4 \, m_\pi^2}~,
\label{gxq}
\ee
where ${\overline {|{\cal M}|^2_X}}$ stands for the square of the
amplitude appropriately averaged/summed over all the variables except for
$\theta_X$ and $q^2$,
\bea
\label{mxq}
&&{\overline {|{\cal M}|^2_X}}=\left|
S\right|^2  - {2\over
3}\left(1-3\cos^2\theta_X\right)
\left(q_0^2-q^2\right)\, \left(1-{4m_\pi^2\over q^2}\right) \, {\rm Re}\left(S
D_1^*\right)
 \\ \nn
&+& {1\over 9}\, \left(1-3\cos^2\theta_X\right)^2 \,
\left(q_0^2-q^2\right)^2\left(1-{4m_\pi^2\over
q^2}\right)^2 \left|D_1 \right|^2
+{8 \over 9}\, \left(q_0^2-q^2\right)^2 \,  \left|D_2 \right|^2 \\ \nn
&-& {8\over
27 }\left(1-3\cos^2\theta_X\right)\, \left(q^2+2q_0^2\right) \,
\left(q_0^2-q^2\right) \, \left(1-{4m_\pi^2\over q^2}\right) \,
{\rm Re}\left(D_2 D_3^* \right) \\ \nn
&+&{8\over 9}\,\left (q^2 -4m_\pi^2 \right )^2\,
\left [ 1+{1\over 3}\, \left( 1 + 3
\cos^2\theta_X\right)\,{q_0^2-q^2\over q^2}+\left(1-3 \cos^2\theta_X \right)^2\,
{(q_0^2-q^2)^2\over 9\,
(q^2)^2}\right] \, \left|\,D_3\right|^2~,
\eea
where $q_0$ stands for the total energy of the two pions in the rest frame of
the initial $\Upsilon$ resonance: $q_0 = (P \cdot q)/\sqrt{P^2}=(M'^2 -
M^2+q^2)/2M'\approx M'-M$ with $M'$ ($M$) standing for the mass of the initial
(final) $\Upsilon$ resonance.

The four form factors in Eq.(\ref{sddd}) can be expressed in terms of the form
factors in the expression (\ref{abc}):
\bea
S &=& \left ( A+ {1 \over 3} \, C \right ) \, (q^2-2 \, m_\pi^2) + \lambda \,
m_\pi^2 + {1 \over 12} \, \left ( B-  {2 \over 3} \, C \right ) \, \left [ 3 \,
q_0^2 - (q_0^2-q^2) \, \left ( 1- {4 m_\pi^2 \over q^2} \right ) \right ] \nn \\
D_1 &=& - {1 \over 4 } \, \left ( B-  {2 \over 3} \, C \right ) ~,  ~~~
D_2 = {1 \over 6} \, C \,  \left ( 1+ {2 m_\pi^2 \over q^2} \right ) ~,~~~
D_3 = - {1 \over 4} \, C~.
\label{sdab}
\eea

It can be noted that the $D_2$ and $D_3$ terms in Eq.(\ref{sdab}) are both
proportional to $C$, thus if $\lambda$ is set at zero and no polarization
information  is being used, the shape of the distribution of the decay rate over
the phase space is determined only by the parameters
\be
\left | C \over A+C/3 \right | = \left | c \over 1+c/3 \right | ~~~{\rm
and}~~~ {B- 2 \, C /3 \over A+ C/3}= {b-2 \, c/3 \over 1+ c/3}~.
\label{params}
\ee
However, if $b$ and $c$ are allowed to be complex numbers, there is a continuous
set of transformations of $b$ and $c$ that leave the quantities (\ref{params})
intact:
\be
c \to {\tilde c} = {3 \, c \, e^{i \phi} \over 3+ c \, (1-e^{i \phi})}~~~{\rm
and}~~~b \to {\tilde b}= {3 \, b - 2\, c \, (1-e^{i\phi}) \over 3+ c \, (1-e^{i
\phi})}~.
\label{trans}
\ee
In case both $b$ and $c$ are constrained to be real, there still remains a
two-fold ambiguity, corresponding to setting $e^{i \phi}=-1$:
\be
c \to {\tilde c} = - {3 \, c  \over 3+ 2 \, c }~~~{\rm and}~~~b \to {\tilde b}=
{3 \, b - 4\, c  \over 3+ 2 \, c }~.
\label{transd}
\ee

The noticed ambiguity can be resolved if a polarization information could be
included in the analysis. This would also provide a more direct access to
fitting the spin-dependent form factor $C$. We illustrate this conclusion by
writing the distribution of the decay rate over the angle $\theta_q$ between the
axis of the initial $e^+$ and $e^-$ beams and the total spatial momentum $\vec
q$ of the dipion:
\be
{d\Gamma\over d\cos\theta_q \, d q} \propto {\overline {|{\cal M}|^2_q}} \,
\sqrt{q_0^2 - q^2} \, \sqrt{q^2-4 \, m_\pi^2}~,
\label{gtq}
\ee
with ${\overline {|{\cal M}|^2_q}}$ now standing for the square of the
amplitude averaged/summed over all the variables except for
$\theta_q$ and $q^2$,
\bea
&& \!\!\!\!\!\!{\overline {|{\cal M}|^2_q}}=\left|
S\right|^2-{2\over
3}\left(1-3\cos^2\theta_q\right)
\left(q_0^2-q^2\right){\rm Re}\left(S D_2^*\right)+{10\over 9}\left(1-{3\over 5}
\cos^2\theta_q \right)
\left(q_0^2-q^2\right)^2 \left|\,D_2\right|^2 \nn \\
&+&{4\over 45}\left(q_0^2-q^2\right)^2\left(1-{4m_\pi^2\over
q^2}\right)^2 \left|\,D_1 \right|^2
\label{mtq}   \\
&-& {4\over
135}\left(1-3\cos^2\theta_q\right)\left(q^2+2q_0^2\right)
\left(q_0^2-q^2\right)\left(1-{4m_\pi^2\over q^2}\right)^2
{\rm Re}\left(D_1 D_3^* \right)\nn \\
&+&{8\over 9}\,\left ( q^2 -4m_\pi^2 \right )^2\,
\left[ 1+{47\over 60}\, \left(1-{21\over
47}\cos^2\theta_q\right) \,{q_0^2-q^2\over q^2}+\left(1-{3\over
5}\cos^2\theta_q \right)\, {(q_0^2-q^2)^2\over 9\,
(q^2)^2}\right ] \, \left|\,D_3\right|^2~.\nn
\eea
One can notice that there is only interference in the angular distribution
between the $S$ and $D_2$ terms as well as between $D_1$ and $D_3$, since the
former two terms correspond to the $S$-wave motion of the pions in their c.m.
frame, while the latter two describe the $D$-wave motion. Clearly, the presence
of the $S D_2$ interference term in the angular distribution should facilitate
an observation of the $C$ form factor (proportional to $D_2$), even if it is
somewhat small in comparison with $A$ and $B$.

The rescattering between the two pions gives rise to phases of the terms in the
amplitude. These phases however depend only on $q^2$ and on the angular momentum
of the pions in their c.m. frame. Thus the terms  $S$ and $D_2$ receive the $S$-
wave scattering phase factor $\exp i\delta_0$ and the terms $D_1$ and $D_3$ get
the $D$-wave factor $\exp i \delta_2$. The $\pi \pi$ scattering phases have been
a subject of many studies (see e.g in Ref.\cite{oop}). In the context of the
present discussion it should be clearly understood that including these phases
implies going beyond the leading soft pion limit, which is likely necessary
anyway for considering the transition $\Upsilon(3S) \to \Upsilon(1S) \, \pi \pi$
with a relatively high energy release. In such situation a description of the
transition amplitude in terms of the partial waves, as in Eq.(\ref{sddd}), is
more consistent than using the form factors defined by the parametrization in
Eq.(\ref{abc}). Indeed, according to Eq.(\ref{sdab}) the terms $D_2$
and $D_3$ are both proportional to the form factor $C$ in Eq.(\ref{abc}). The
$\pi \pi$ rescattering however modifies in a different way the terms $D_2$ and
$D_3$, so that it would be impossible to describe them in terms of one and the
same form factor $C$. In other words, the parametrization in Eq.(\ref{abc}) is
only suitable in the low energy limit. Going beyond this limit requires a more
general parametrization of the amplitude. On the other hand the expression in
Eq.(\ref{sddd}) with the coefficients being functions of $q^2$ is a general
expression for the $S$-wave and all possible types of $D$-wave motion, and is
limited only by the absence of higher partial waves.

It can be mentioned in connection with the discussion of the complex phases,
that the distribution described by the equations (\ref{gtq}) and (\ref{mtq}) is
in fact not sensitive to the phase shifts due to the $\pi \pi$ rescattering, so
that the interfering terms in Eq.(\ref{mtq}) are relatively real. Therefore an
observation of a deviation from this behavior would signal either a presence of
higher partial waves, or complex phase effects from rescattering in the exotic
$\Upsilon \pi$ channel. An explanation of the peculiar spectrum in the decay
$\Upsilon(3S) \to \Upsilon(1S) \, \pi \pi$ based on four-quark states has been
suggested\cite{mv2,absz,gscp} in the past, but has never had a phenomenological
success. It would thus be most interesting if a nontrivial dynamics in this
channel manifested itself in such a subtle effect.

Finally, a remark is due on the relative significance of the quarkonium spin
dependent part of the amplitude, described by the form factor $C$ in
Eq.(\ref{abc}) or by $D_2$ and $D_3$ in Eq.(\ref{sddd}). This term is naturally
suppressed by the inverse of the heavy quark mass, and normally should be small
in comparison with the dominant $S$ term. This appears to be indeed the case for
the `well behaved' dipion transitions. However it is not necessarily the case
for the $\Upsilon(3S) \to \Upsilon(1S) \, \pi \pi$ decay. Indeed, one may argue
that the overall rate of this transition is suppressed, so that in fact the $S$
term is small and does not dominate over the spin-dependent part\cite{mv1}.
Clearly, a definite input from experiment would be extremely helpful in
exploring this possibility. As discussed previously in this paper, a study of
the polarization-correlated angular distributions, e.g. of the one described by
Eqs. (\ref{gtq}) and (\ref{mtq}), would provide a more direct access to the spin
dependent terms.

\section*{Acknowledgments}
The authors are thankful to R.S. Galik for illuminating correspondence regarding
Ref.\cite{cleo}. The work of MBV is supported in part by the DOE grant
DE-FG02-94ER40823.

\end{document}